\documentclass[twocolumn]{aastex63}
\usepackage{hyperref}
\usepackage{color}
\usepackage{comment}
\usepackage{ifthen}
\usepackage{grffile}
\usepackage[utf8]{inputenc}
\usepackage{bm}
\usepackage{mathrsfs}
\usepackage{enumitem}
\usepackage{amsmath,mathtools}

\begin{document}

\defcitealias{Paine:2022fo}{P22}
\defcitealias{Darling:2023ao}{D23}
\defcitealias{Hosek:2025kf}{H25}

\title{IRS 9: The Case for a Dynamically-Ejected Star from the Galactic Center}

\author[0000-0003-2874-1196]{Matthew W. Hosek Jr.}
\correspondingauthor{Matthew W. Hosek Jr.}
\altaffiliation{Brinson Prize Fellow}
\affiliation{UCLA Department of Physics and Astronomy, Los Angeles, CA 90095}
\email{mwhosek@astro.ucla.edu}

\author[0000-0001-9554-6062]{Tuan Do}
\affiliation{UCLA Department of Physics and Astronomy, Los Angeles, CA 90095}

\author[0000-0002-9802-9279]{Smadar Naoz}
\affiliation{UCLA Department of Physics and Astronomy, Los Angeles, CA 90095}
\affiliation{Mani L. Bhaumik Institute for Theoretical Physics, Department of Physics and Astronomy, UCLA, Los Angeles, CA 90095, USA}

\author[0000-0003-0984-4456]{Sanaea C. Rose}
\affiliation{Center for Interdisciplinary Exploration and Research in Astrophysics (CIERA), Northwestern University, 1800 Sherman Ave, Evanston, IL 60201}

\author[0000-0002-7476-2521]{Gregory D. Martinez}
\affiliation{UCLA Department of Physics and Astronomy, Los Angeles, CA 90095}

\author[0000-0003-3230-5055]{Andrea M. Ghez}
\affiliation{UCLA Department of Physics and Astronomy, Los Angeles, CA 90095}

\author[0000-0001-7003-0588]{Rebecca Lewis-Merrill}
\affiliation{UCLA Department of Physics and Astronomy, Los Angeles, CA 90095}

\author[0000-0001-9611-0009]{Jessica R. Lu}
\affiliation{Department of Astronomy, 501 Campbell Hall, University of California, Berkeley, CA, 94720}

\author[0000-0001-5972-663X]{Shoko Sakai}
\affiliation{UCLA Department of Physics and Astronomy, Los Angeles, CA 90095}

\author[0000-0003-2861-3995]{Jay Anderson}
\affiliation{Space Telescope Science Institute, 3700 San Martin Drive, Baltimore, MD 21218, USA}

\begin{abstract}
Measuring stellar motions at the Milky Way's Galactic center (GC)
provides unique insight into the dynamical processes within galactic nuclei.
We present proper motion measurements for 23 SiO-maser emitting stars within 45'' of SgrA*,
including four previously reported to have velocities exceeding their local escape velocities (i.e., they are ``locally unbound'' from the GC).
Derived from 14 epochs of HST WFC3-IR observations (2010 -- 2023), our measurements have a median precision of 0.038 mas yr$^{-1}$ -- up to
$\sim$100x more precise then previous constraints for some sources.
By combining these proper motions with published radial velocities, we derive updated 3D velocities for the masers
and find that only one is locally unbound (IRS 9; $v_{3d}$ = 370 $\pm$ 1.2 km s$^{-1}$).
Orbit integrations place the first constraints on the orbit of IRS 9, which is bound to the GC at
larger radii with $r_{peri}$ $\geq$ 0.100 $\pm$ 0.005 pc and $r_{apo}$ $\geq$ 5.25 $\pm$ 0.18 pc.
IRS 9's high velocity relative to stars at similar radii in the Nuclear Star Cluster
makes it a candidate to have experienced a strong dynamical interaction in order to place it on its orbit.
We explore the Hills mechanism as a possible origin, but binary evaporation and ejection velocity limits indicate
that IRS 9 is unlikely to have experienced such an event in the past 0.4 Myr (the timescale constrained by the orbit integrations).
Alternative mechanisms that could produce IRS 9 include binary supernova disruption, two-body interactions, and stellar collisions.
Identifying additional stars like IRS 9 will be essential for understanding these various dynamical processes.
\end{abstract}


\section{Introduction}
Due to its proximity, the Milky Way's Galactic center (GC) is the only
galactic nucleus for which it is possible to study the kinematics
of individual sources in great detail.
As such, the GC provides a window into the stellar populations
within a galactic nucleus and the dynamical processes that act upon them.
Measurements of stellar orbits at the GC
have proven the existence of a central supermassive black hole (SMBH) with a
mass of $\sim$4 x10$^6$ M$_{\odot}$ associated with the emissive source SgrA*
\citep[e.g.][]{Schodel:2002qq, Ghez:2003ul, Ghez:2008tg, Gillessen:2009uo, Do:2019gr, GRAVITY-Collaboration:2022kk}.
Surrounding SgrA* is the Nuclear Star Cluster (NSC),
the densest and most massive star cluster in the Galaxy,
which is composed of primarily old stars ($\gtrsim$5 Gyr) with a total mass of $\sim$10$^7$ M$_{\odot}$
within R $\lesssim$ 5 pc \citep[e.g.][]{Schodel:2014zl, Chatzopoulos:2015lq, Schodel:2020pp, Gallego-Cano:2020fx, Chen:2023os}.
Similar components have been found in the nuclei of many galaxies beyond the Milky Way \citep[e.g.][]{Kormendy:2013bs, Neumayer:2020nk, Fahrion:2024ug},
and so the GC offers a template for understanding these complex systems.

SiO masers provide useful probes of stellar kinematics of this region.
A component of the NSC population, stars with SiO maser emission are evolved red giant or
supergiant stars with extended circumstellar envelopes
that produce emission lines at radio wavelengths associated with
SiO molecules \citep[e.g.][]{Reid:2002uv, Kemball:2007dr}.
Because of their value in defining an astrometric reference frame
for studies of stellar orbits at the GC \citep[e.g.][]{Menten:1997vh, Ghez:2008tg, Plewa:2015ud, Sakai:2019fm},
these sources have been the target of long-running radio campaigns to measure
their proper motions and radial velocities to high precision \citep{Reid:2003ai, Reid:2007jk, Li:2010jf, Paine:2022fo, Darling:2023ao, Tsuboi:2025qv}.

Interestingly, four of the 25 known
stellar masers near SgrA* ($r_{2d}$ $\leq$ 45'', where $r_{2d}$ is the projected radius on the sky)
have been reported to be ``locally unbound'', i.e., they exhibit velocities that
exceed the escape velocities at their projected radii from SgrA* according to GC mass models.
The star IRS 9 was the first maser source identified
to be locally unbound on the basis of its large radial velocity \citep{Reid:2007jk}.
Three additional masers were identified as being locally unbound by  \citet[][hereafter P22]{Paine:2022fo}
on the basis of their large proper motions (SiO-16, SiO-21, and SiO-25).
The presence of so many high-velocity sources within this limited sample
raises the question of what dynamical mechanism(s) might be responsible for
producing them.

In this paper, we report proper motion measurements for 23 of the
stellar SiO masers near SgrA*, including the four proposed high-velocity sources,
derived using multi-epoch Hubble Space Telescope (HST) observations of the GC
(14 epochs between 2010 -- 2023).
These proper motions are combined with radial velocities in the literature
to calculate updated 3D velocities for the masers ($\mathsection$\ref{sec:obs}),
which are then compared to the GC escape velocity curve to identify sources
that are locally unbound ($\mathsection$\ref{sec:masers_3d}).
We then use orbit integrations to place the first constraints on the orbit
of IRS 9, which is found to be
bound to the GC at larger radii
($\mathsection$\ref{sec:irs9_orb}).
The velocity and orbital eccentricity of IRS 9
is compared to the expected distributions for stars in the NSC ($\mathsection$\ref{sec:irs9_origin}),
and dynamical mechanisms that might have accelerated IRS 9 on
its current orbit are explored ($\mathsection$\ref{sec:mechanisms}).
Finally, we summarize our findings in $\mathsection$\ref{sec:conclusions}.

\section{Observations and Measurements}
\label{sec:obs}
\subsection{HST Astrometry and Proper Motions}
The HST observations and procedure for extracting astrometric measurements
used in this work are described in \citep[][hereafter \citetalias{Hosek:2025kf}]{Hosek:2025kf}\footnote{All the HST data used in this paper can be found in MAST: \dataset[doi:10.17909/vre6-c497]{https://doi.org/10.17909/vre6-c497}}.
Briefly, 14 epochs of WFC3-IR observations of a 2' x 2' field centered on SgrA* ($\alpha$(J2000) = 17$^h$45$^m$40.04$^s$, $\delta$(J2000) = -29$^{\circ}$00$'$28$''$.10)
were obtained using the F153M bandpass filter between 2010.6261 -- 2023.6178 (see Table 1 from \citetalias{Hosek:2025kf}).
Stellar photometry and astrometry is extracted in a two-step process.
First, the \texttt{FORTRAN} routine \texttt{img2xym\_wfc3ir} \citep[a precursor to the package \texttt{hst1pass} described in][]{Anderson:2022vs}
is used for an initial round of star detection in each image using a library of spatially-variable point-spread functions (PSFs)
for the WFC3-IR F153M filter.
For each image, a perturbation is applied to the PSFs to minimize the PSF residuals
and a first-order polynomial transformation is used to align the image into a common
reference frame defined for the epoch.
Second, the \texttt{FORTRAN} routine \texttt{KS2} \citep[][]{Anderson:2008qy, Bellini:2017xy, Bellini:2018ow}
is used to combine the individual images into a combined image for the epoch, run source detection on that combined image,
and then extract measurements for the detected sources back in the individual images.
This allows for significantly deeper source detection than is possible in the first step alone.

This process produces a starlist with the astrometric and photometric measurements
for each epoch.
The astrometric uncertainty of a given star is calculated as the error-on-the-mean of its
measured positions across the individual images within the epoch ($\sigma_{HST}$ = $\sigma_{img}$ / $\sqrt{N_{img}}$,
where $\sigma_{HST}$ is the astrometric error and $\sigma_{img}$ is the standard deviation of the positions across $N_{img}$).
The photometric uncertainty is calculated as the standard deviation of the instrumental magnitudes
measured across the images within the epoch \citep[see][]{Hosek:2015cs}.
Instrumental magnitudes are transformed into Vega magnitudes using the \texttt{KS2}
zeropoints from \citet{Hosek:2018lr}.
A typical starlist in the dataset contains $\sim$50,000 stars and extends to a depth of
F153M $\sim$ 22.5 mag (the 95th percentile of the extracted magnitudes).

Proper motions are derived as described in \citetalias{Hosek:2025kf}.
First, the HST astrometry for each epoch is transformed into the Gaia-CRF3
reference frame \citep{Gaia-Collaboration:2022cm}, which is tied
to the International Celestial Reference System \citep[ICRS;][]{Arias:1995rb}.
Stars are then cross-matched between epochs and proper motions
are calculated using a method that uses Gaussian Processes to simultaneously
model systematic correlations in the astrometry (see section 3.2 of \citetalias{Hosek:2025kf}).
Finally, the proper motions are transformed into a SgrA*-at-Rest frame
(a reference frame where SgrA* is at the position origin and is at rest)
using the ICRS position and proper motion of SgrA* from \citet{Xu:2022sx}.
Overall, the astrometric transformations between these reference frames incur a
systematic error of (0.025, 0.019) mas yr$^{-1}$ in the ($\mu_{\alpha*}$ and $\mu_{\delta}$) proper motions
and (0.213, 0.471) mas in the ($\alpha^*$, $\delta$) positions\footnote{Note that $\alpha^*$ is used to designate the
RA coordinate projected on a tangent plane on the sky, i.e. RA cos(DEC), while $\delta$ is used to designate the DEC coordinate.}.
These systematic errors are added in quadrature with the measurement errors to get the final
proper motion errors. The best-measured stars in the dataset achieve
proper motion errors of $\sim$0.03 mas yr$^{-1}$ \citepalias{Hosek:2025kf}.

\begin{figure}
\begin{center}
\includegraphics[scale=0.38]{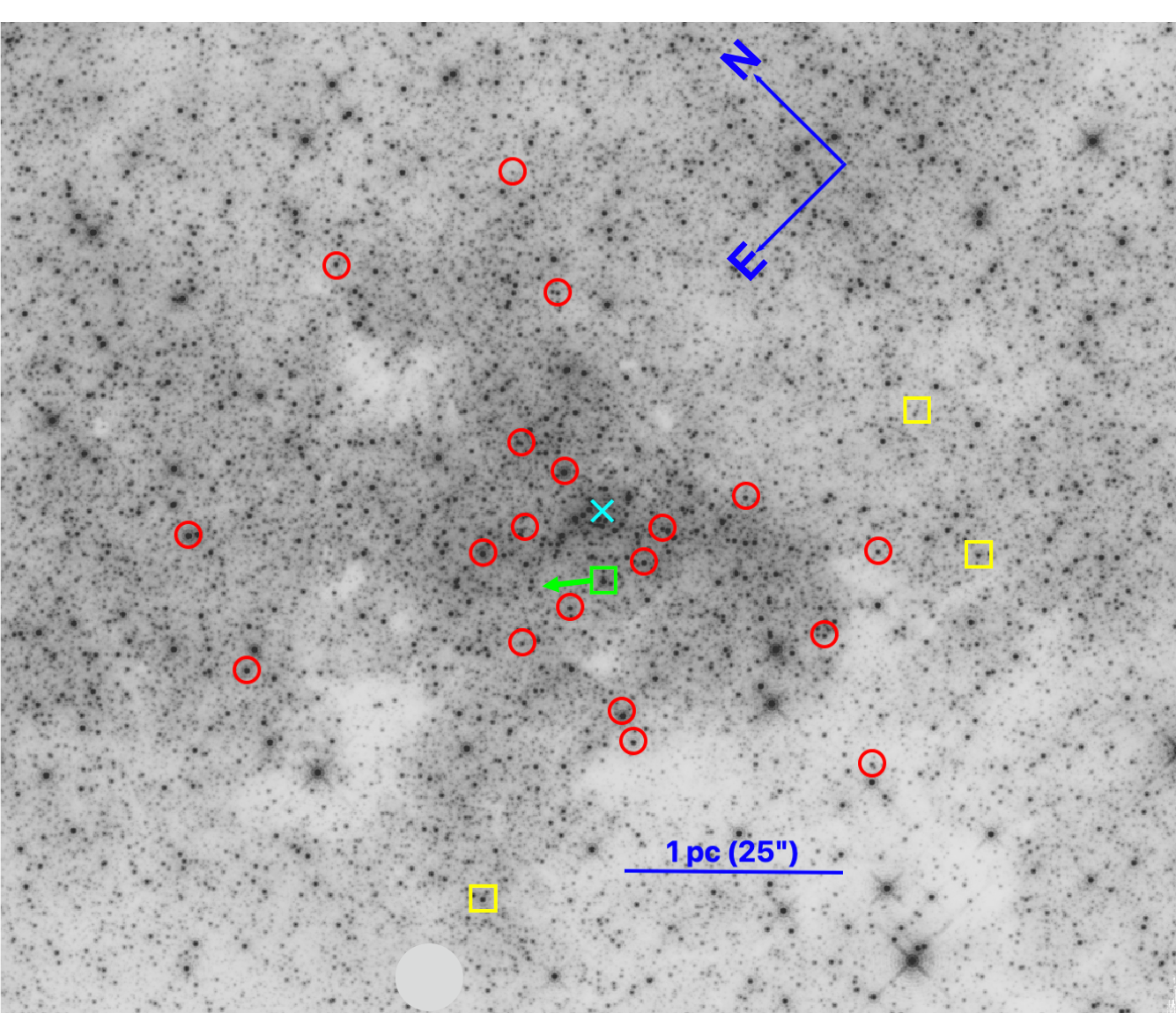}
\caption{The 23 SiO masers examined in this study, overlaid on an HST F153M image. The green square
represents IRS 9, which is found to exhibit a velocity larger than the maximum
escape velocity at its radius (i.e., it is locally unbound from the GC).
The green arrow shows the direction of its HST proper motion (not to scale).
The yellow squares correspond to three masers proposed to be locally unbound
(SiO-16, SiO-21, and SiO-25),
but the new HST proper motions indicate that they do not exceed their maximum escape velocities.
The red circles show the positions of the other masers in the sample.
All sources have projected distances of R $\leq$ 45'' ($\sim$1.8 pc) from SgrA* (cyan x).
}
\label{fig:img}
\end{center}
\end{figure}

\subsection{Identifying Masers and Calculating Their 3D Velocities}
\label{sec:maser_pms_text}

We identify HST counterparts to a sample of 25 SiO masers within $r_{2d}$ $\leq$ 45'' ($r_{2d}$ $\lesssim$ 1.8 pc) of SgrA*
that have radio-based proper motion measurements \citep[\citetalias{Paine:2022fo}, see also][hereafter D23]{Darling:2023ao}.
The radio proper motions are used to calculate the expected positions of the masers
at a time of 2015.5, and the HST proper motion are used to do the same for the HST sources.
An HST source is matched with a maser if its position is consistent with the radio position
to within a matching radius of 0.15$''$ and if the reduced chi-squared statistic of the HST kinematic motion model
fit ($\chi^2_{red}$) is less than 5 (a conservative quality cut to eliminate poorly fit HST sources, see \citetalias{Hosek:2025kf}).
This results in successful matches for 23 of the 25 masers, including all 4 proposed
locally unbound sources (Figure \ref{fig:img}).
The two unmatched sources are SiO-12, for which no HST source was found within
the matching radius,
and SiO-17, which did have an HST source within the matching radius but with a poor
HST kinematic motion model fit ($\chi^2_{red}$ $>$ 5), likely due to the influence of a nearby bright star
biasing the astrometry.

The 3D velocities of the masers ($v_{3d}$) are calculated by combining their HST proper motions
with the radial velocities from \citetalias{Paine:2022fo}\footnote{We use the radial velocities from \citetalias{Paine:2022fo}
because they report uncertainties on the radial velocities while \citetalias{Darling:2023ao} do not.
Regardless, the \citetalias{Paine:2022fo} and \citetalias{Darling:2023ao} radial velocities are consistent within 0.4 km/s,
which is well within the uncertainties of the 3D velocities.}.
These radial velocities are reported in the local standard of rest (LSR),
which we assume to be equivalent to their radial velocity relative to SgrA*
\citep[i.e., SgrA* has no significant line-of-sight motion with respect to the LSR
and is at rest relative to the dynamical center of the Milky Way;][]{Reid:2004xh, Reid:2020jo}.
To convert the proper motions from angular to physical units, we assume a distance
to SgrA* of 8.1 kpc  \citep[the average of recent distance estimates;][]{Do:2019gr, GRAVITY-Collaboration:2022kk}.

A catalog of the masers with HST proper motions and updated 3D velocities is
provided in Appendix \ref{app:pm_cat}.
The masers vary in brightness between 10.95 mag $\leq$ F153M $\leq$ 17.73 mag
and their proper motion measurements have a median uncertainty of 0.038 mas yr$^{-1}$.
For the masers closest to SgrA* ($r_{2d}$ $\leq$ 20", or $r_{2d}$ $\lesssim$ 0.8 pc), the HST measurements achieve similar
uncertainties as previous radio-based studies,
as these sources have been
extensively monitored via radio observations \citep{Reid:2003ai, Reid:2007jk, Li:2010jf, Paine:2022fo, Darling:2023ao}.
For sources at larger radii ($r_{2d}$ $\geq$ 20", or $r_{2d}$ $\gtrsim$ 0.8 pc)), the HST measurements
improve the uncertainties by as much as $\sim$100x relative to radio-based measurements.
This is because there is often significantly more HST astrometry available for the
sources at large radii compared to radio studies \citep[e.g.][]{Paine:2022fo}.
A detailed comparison between the HST and radio proper motion
measurements for the masers is provided Appendix \ref{sec:maser_comp}.

\section{Results}
\label{sec:results}

\subsection{Are the Masers Locally Unbound From the GC?}
\label{sec:masers_3d}

We compare the 3D velocities of the masers to the escape velocities
at their projected radii ($r_{2d}$) to determine if they are locally unbound from the GC.
Since the line-of-sight distances of the masers are not well constrained,
$r_{2d}$ represents the minimum possible physical radius of a given source from SgrA*.
Thus, the escape velocity at $r_{2d}$ is the maximum possible
escape velocity for that source ($v_{esc, max}$).
A maser is locally unbound if it fulfills the condition:
\begin{equation}
\label{eq:enc_mass}
v_{3d} \geq v_{esc, max}
\end{equation}
\begin{equation*}
v_{esc, max} = \sqrt{2GM_{enc} / r_{2d}} \, ,
\end{equation*}
where $v_{3d}$ is the 3D velocity of the maser,
$M_{enc}$ is the enclosed mass within $r_{2d}$,
and $G$ is the gravitational constant \citep[e.g.][]{Reid:2003ai}.

To calculate $M_{enc}$, we adopt a mass model for the GC
with two components: the supermassive black hole (SMBH),
which dominates the gravitational potential for r $\lesssim$ 1pc,
and the Nuclear Star Cluster (NSC), which dominates
between 1 pc $\lesssim$ r $\lesssim$ 30 pc \citep[e.g.][]{Launhardt:2002hl, Sormani:2020my}.
We adopt a SMBH mass of (4.14 $\pm$ 0.16) x 10$^6$ M${_{\odot}}$,
which represents the average and standard deviation of the masses reported by \citet{Do:2019gr}
and \citet{GRAVITY-Collaboration:2022kk}.
For the NSC, we adopt the mass density distribution from the
best-fit axisymmetric model from \citet[][see their Equations 17 and 19]{Chatzopoulos:2015lq}:
\begin{equation}
\rho(R, z) = \frac{3 - \gamma}{4 \pi q} \frac{a_{NSC} M_{NSC}}{m^\gamma (m + a_{NSC})^{4 - \gamma}} \, ,
\end{equation}
where
\begin{equation}
m^2 = R^2 + \frac{z^2}{q^2} \, ,
\end{equation}
and $R$ = $\sqrt{x^2 + y^2}$, $q$ = 0.73 $\pm$ 0.04, $\gamma$ = 0.71 $\pm$ 0.12, $a_{NSC}$ = 5.9 $\pm$ 1.07 pc,
and $M_{NSC}$ = (6.1 $\pm$ 0.3) * 10$^7$ M$_{\odot}$.

At a given $r_{2d}$, the uncertainty in $v_{esc, max}$ is calculated via a
Monte-Carlo simulation.
We draw 100 samples of the SMBH mass and NSC model parameters, each
perturbed by a random amount drawn from a Gaussian distribution with a width equal to its
uncertainty, and then calculate $v_{esc, max}$ for each sample.
The uncertainty in $v_{esc, max}$ is
calculated as the standard deviation of the values across the samples.

\begin{figure}
\includegraphics[scale=0.35]{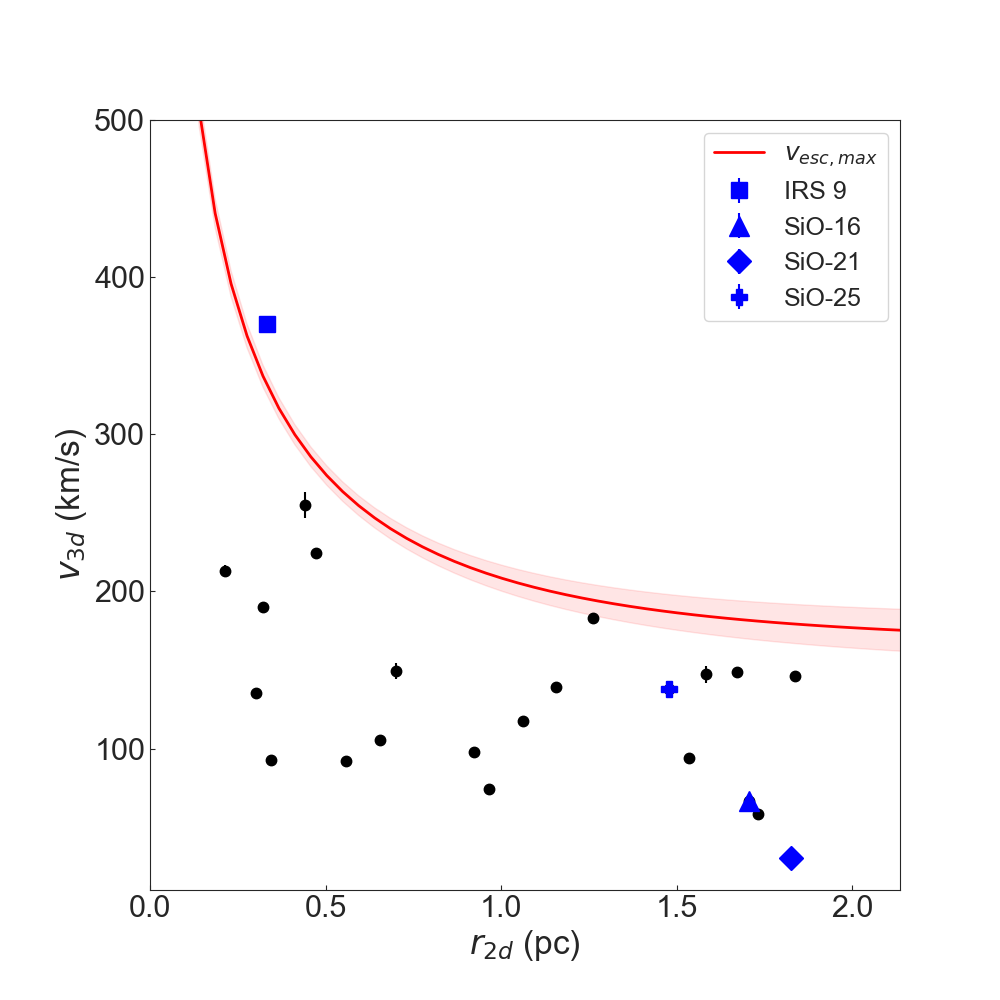}
\caption{Maser $v_{3d}$ as a function of $r_{2d}$.
A model for $v_{esc, max}$ and the corresponding 1$\sigma$ uncertainty in that model
is shown by the red line and shaded region.
Previously reported masers with $v_{3d} > v_{esc, max}$ are highlighted by
blue markers: IRS 9 (blue square), SiO-16 (blue triangle), SiO-21 (blue diamond),
and SiO-25 (blue plus sign).
When the HST proper motions are used, only IRS 9 has $v_{3d} > v_{esc, max}$.}
\label{fig:maser_3d}
\end{figure}

We find that only IRS 9 has a 3D velocity
that is significantly above $v_{esc, max}$ at its projected radius (Figure \ref{fig:maser_3d}).
At $r_{2d}$ = 0.33 pc, IRS 9 exhibits $v_{3d}$ = 370.4 $\pm$ 1.2 km s$^{-1}$
compared to $v_{esc, max}$ = 331.1 $\pm$ 6.5 km s$^{-1}$.
This corresponds to a difference at the 5.9$\sigma$ significance level
(where $\sigma$ is the combined uncertainty in $v_{3d}$ and $v_{esc, max}$),
indicating that IRS 9 is locally unbound.

However, the remaining masers in the HST sample have 3D velocities below their $v_{esc, max}$ limits,
including the proposed high-velocity sources SiO-16, SiO-21, and SiO-25 (Figure \ref{fig:maser_3d}).
The updated $v_{3d}$ values for these masers
are over $\sim$1000 km s$^{-1}$ lower than those reported in \citetalias{Paine:2022fo},
representing the largest discrepancies between the HST and radio
measurements by far (Appendix \ref{sec:maser_comp}).
\citetalias{Paine:2022fo} note that the radio-based proper motions for these sources are derived from
very limited radio astrometry (only 2 epochs over a $\sim$2 year time baseline),
and so the HST proper motion measurements are likely to be more reliable.

For later discussion of possible dynamical mechanisms for IRS 9 ($\mathsection$\ref{sec:mechanisms}),
it is useful to convert the star's observed velocity into the velocity at infinity
relative to the SMBH ($v_{\infty, smbh}$; the velocity at infinity if only
the SMBH is considered for the gravitational potential).
Assuming the minimum possible physical radius of IRS 9 (i.e., its physical radius is equal to its projected radius $r_{2d}$ = 0.33 pc),
then the observed velocity corresponds to $v_{\infty, smbh}$ = 134 $\pm$ 18 km s$^{-1}$.
This represents a lower limit for $v_{\infty, smbh}$,
because if the true physical radius of IRS 9 is larger, then
the corresponding $v_{\infty, smbh}$ would also be larger (Appendix \ref{app:los_dist}).

It should be noted that the 3D velocity of IRS 9 is dominated by its
large radial velocity of -341 $\pm$ 1.2 km s$^{-1}$ \citepalias{Paine:2022fo}.
There is no evidence of significant radial velocity variations for IRS 9
across across many epochs of radio observations \citep[][\citetalias{Paine:2022fo}]{Reid:2007jk},
and independent near-infrared spectroscopy of the star yields similar radial velocity values \citep{Zhu:2008hc}.
This suggests that the radial velocity we adopt for IRS 9 is reasonable.

\subsection{Constraints on the Orbit of IRS 9}
\label{sec:irs9_orb}
While IRS 9 is locally unbound,
it is not necessarily destined to escape the GC.
This is because of the extended mass distribution in the region;
as the radius of IRS 9 increases, then the amount of enclosed mass
also increases. Thus, it is possible for IRS 9 to become bound at a larger
radius and follow an orbit rather than a hyperbolic trajectory.

We use \texttt{galpy} \citep{Bovy:2015bd} to integrate the orbit of IRS 9
in the GC gravitational potential
in order to place the first constraints on its periapse, apoapse, and orbital eccentricity.
The gravitational potential is taken to be a multi-component model containing the
SMBH, NSC, and Nuclear Stellar Disk (which dominates for r $\gtrsim$ 30 pc) derived via
axisymmetric Jeans modeling of the region \citep[Model 2 from][]{Sormani:2020my}.
For IRS 9, 5 out of 6 phase space parameters are known,
with measurements of its 3D velocity and the 2D projected position on the sky.
However, the star's distance along the line-of-sight not well constrained (d$_{los}$, which we define
as distance along the line-of-sight relative to SgrA*).
For these calculations we will assume that $d_{los}$ = 0 pc (i.e., the projected radius is equal to the physical radius),
which produces the ``tightest'' possible orbit that IRS 9 can have.
In other words, constraints on the periapse and apoapse
of the $d_{los}$ = 0 pc orbit represent lower limits of their true values (see Appendix \ref{app:los_dist}).

\begin{figure*}
\centering
\includegraphics[scale=0.35]{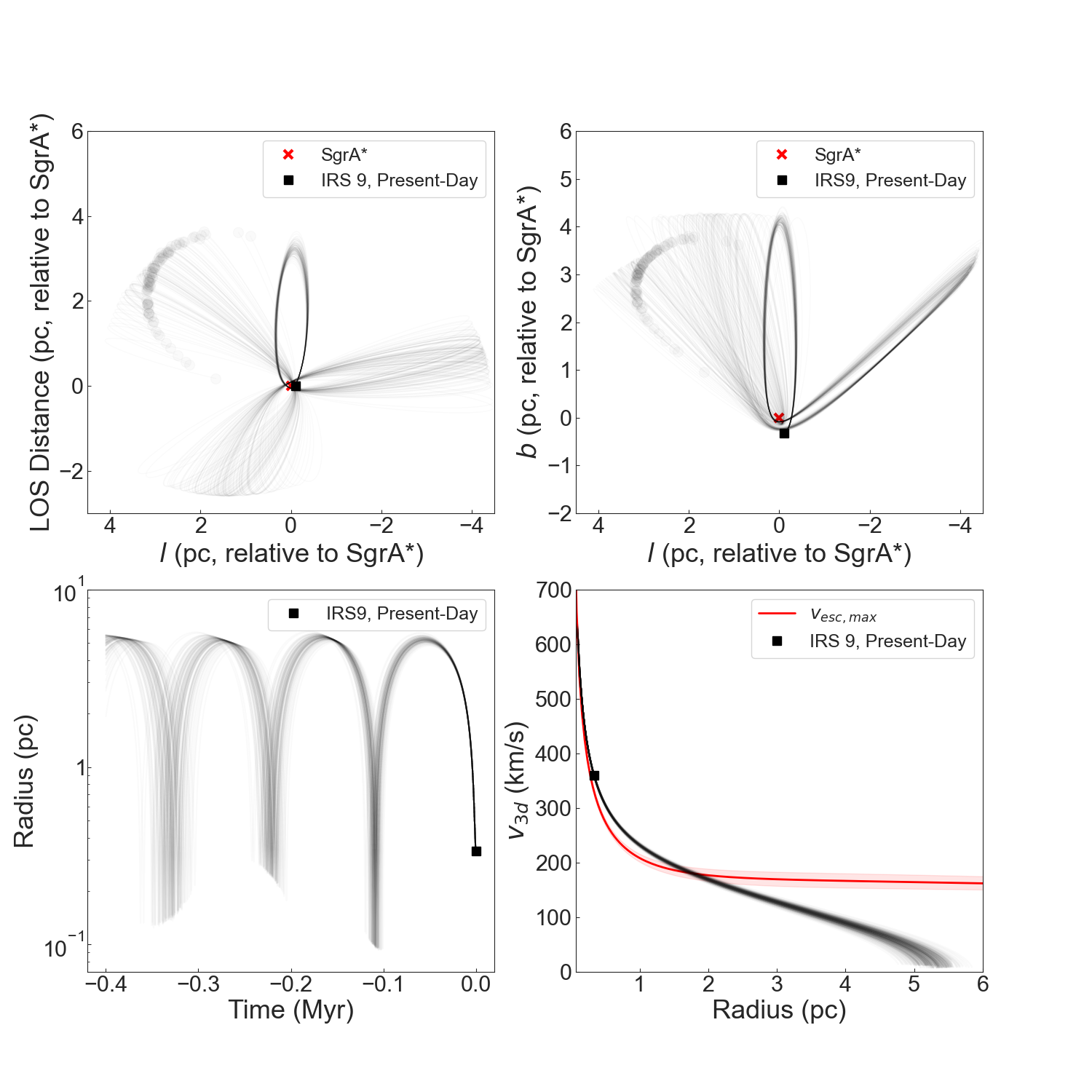}
\caption{The orbit of IRS 9, integrated for the past 0.4 Myr and assuming d$_{los}$ = 0 pc.
In each panel, the black lines represent orbits drawn from an MC simulation over the measured properties of IRS 9, and so the spread
of lines reveals the uncertainty in the orbit.
Top left: A ``top-down'' view of the orbit from the North Galactic Pole (with Earth at negative LOS distance).
The solid black square represents the current position of IRS 9
while the faded black circles represent the end point of each orbit (i.e., 0.4 Myr in the past). SgrA* is represented by the red ``x''.
Top right: An ``edge-on'' view of the orbit from the Galactic Plane, constructed in the same manner as the top-left panel.
Bottom left: The distance of IRS 9 from SgrA* as a function of time.
Bottom right: The $v_{3d}$ of IRS 9 as a function of radius, compared to $v_{esc, max}$ (red line with uncertainty shown as shaded region).
IRS 9 is locally unbound when $v_{3d}$ $>$ $v_{esc, max}$, as is the case at the present-day (at $r_{2d}$ = 0.33 pc). However, the star is
bound to the GC at larger radii as $v_{3d}$ drops below $v_{esc, max}$.
}
\label{fig:irs9_orb}
\end{figure*}

To quantify the impact of measurement uncertainties on the orbit constraints,
we resample the projected position
and 3D velocity of IRS 9 for 500 iterations,
each time using new values drawn from
Gaussian distributions with means and standard
deviations equal to the different measurements and their
corresponding uncertainties.
An orbit is calculated for each iteration that
extends 0.4 Myr into the past,
at which point the orbit constraints
become weak due to the growing impact of
the measurement uncertainties over time
(Figure \ref{fig:irs9_orb}).
This timescale captures the three previous orbits of IRS 9,
which has an average orbital period (i.e., the time between successive
periapse or apoapse passages) of 0.11 $\pm$ 0.005 Myr across the iterations.

The resulting orbits reveal that IRS 9 must be on a highly eccentric orbit
if $d_{los}$ = 0.
For the three most recent orbits captured in this analysis,
the periapse distances ($r_{peri}$)
are 0.100 $\pm$ 0.005 pc, 0.25 $\pm$ 0.02 pc, and 0.14 $\pm$ 0.02 pc,
in order of most recent to furthest back in time.
The corresponding apoapse distances ($r_{apo}$) are
5.25 $\pm$ 0.18 pc, 5.42 $\pm$ 0.16 pc, and 5.24 $\pm$ 0.20 pc.
So, the eccentricities of the orbits ($e = (r_{apo} - r_{peri}) / (r_{apo} + r_{peri})$)
are 0.96 $\pm$ 0.001, 0.91 $\pm$ 0.006, and 0.95 $\pm$ 0.008.

We reiterate that these values for $r_{peri}$ and $r_{apo}$
are lower limits as they increase if $|d_{los}|$ increases (Appendix \ref{app:los_dist}).
Thus, for purposes of exploring possible dynamical mechanisms
for IRS 9 ($\mathsection$\ref{sec:mechanisms}), we conclude
that IRS 9 could not have come closer than 0.100 $\pm$ 0.005 pc to
SgrA* in the past 0.4 Myr.

\section{Discussion}
\label{sec:discussion}

\subsection{Are the Kinematics of IRS 9 Unusual for a Star in the NSC?}
\label{sec:irs9_origin}
To explore the dynamical origin of IRS 9, we first
evaluate if its kinematics are unusual for a star in the NSC.
We define the probability of observing a star with a
velocity greater than or equal to a given $v_{3d, i}$ as:

\begin{equation}
P_{NSC} = \int_{v_{3d, i}}^{\infty} D(v_{3d}, r) dv_{3d} \, ,
\end{equation}

where $D(v_{3d}, r)$ is the probability distribution of $v_{3d}$
at the radius $r$ of the star within the NSC.
$D(v_{3d}, r)$ is assumed to follow a Maxwellian distribution with
a scale parameter equal to $\sigma_{1d}(r)$
(the 1D velocity dispersion at $r$),
as is the case for a stellar isothermal sphere \citep[e.g.][]{Binney:2008bh}.

\begin{figure}
\begin{center}
\includegraphics[scale=0.35]{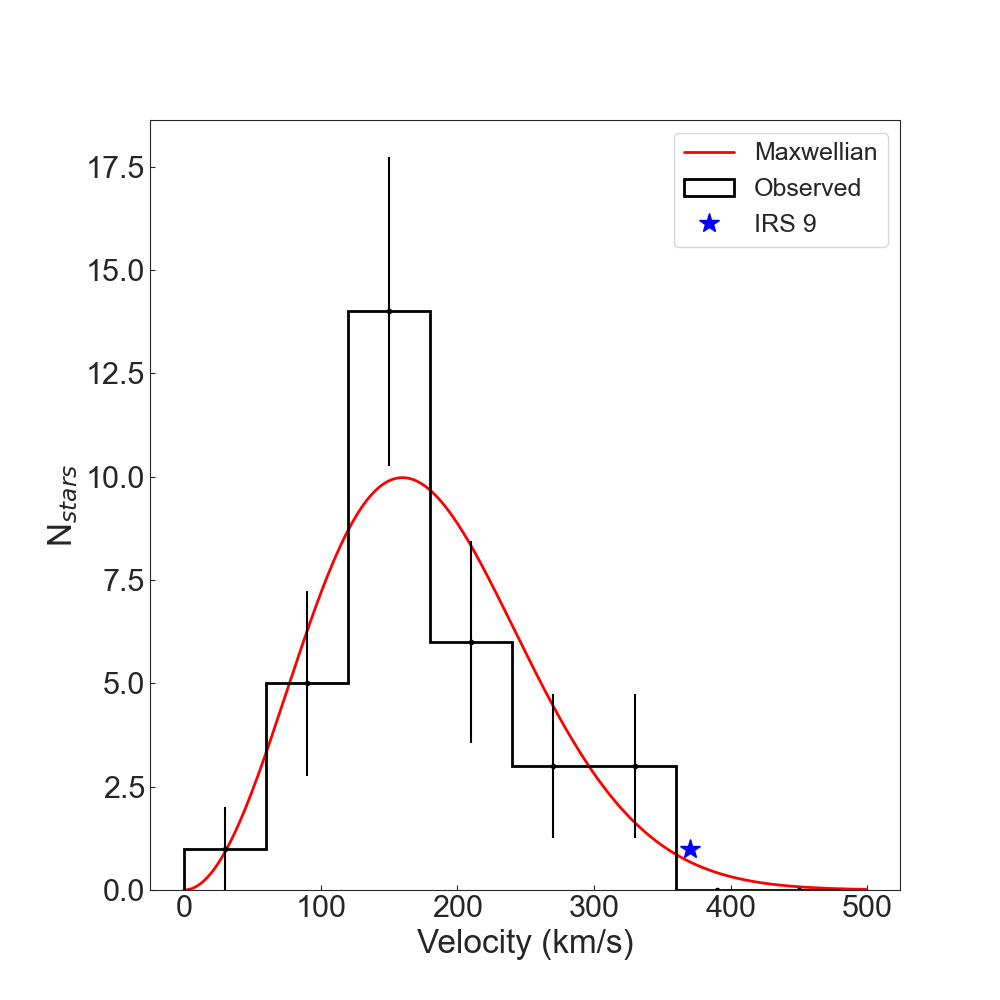}
\caption{The observed $v3d$ distribution of 32 stars with projected radii within $\pm 1"$ of IRS 9
(black histogram), compared to the  $v3d$ of IRS 9 itself (blue star).
The expected $v3d$ distribution of the NSC
is taken to be a Maxwellian distribution
with a scale parameter equal to the velocity dispersion at this radius
derived from a dynamical model of the NSC \citep[][red line]{Chatzopoulos:2015lq}.
From this distribution, we calculate that the probability of observing a star
with a $v3d$ greater or equal to IRS 9 is $\sim$1\%.
}
\label{fig:vel_dist}
\end{center}
\end{figure}

To calculate $P_{NSC}$ for IRS 9, we assume that
$\sigma_{1d}$ = 113 km s$^{-1}$ at its radius based on the
best-fit dynamical models of the NSC from \citet{Chatzopoulos:2015lq}.
The corresponding Maxwellian distribution
provides a reasonable match for the observed $v3d$
distribution for a sample of
32 stars with radii within $\pm 1"$ of IRS 9's projected radius
($7.5'' \leq r_{2d} \leq 9.5''$, corresponding to 0.29 pc $\lesssim$ $r_{2d}$ $\lesssim$ 0.37 pc),
excluding IRS 9 itself (Figure \ref{fig:vel_dist}).
Stellar proper motions and radial velocities are obtained
from the catalogs of \citetalias{Hosek:2025kf} and \citet{Feldmeier-Krause:2017qy},
respectively.
These stars exhibit stellar colors consistent with stars found near the
GC (F127M - F153M $>$ 2.6 mag, e.g. \citetalias{Hosek:2025kf}).

Under these assumptions, we calculate that P$_{NSC}$ $\sim$ 1\%
to observe a star with $v3d$ equal to that of IRS 9 at its projected radius.
While this does not conclusively identify IRS 9 as a velocity outlier
within the NSC, it shows that it is a candidate to have
experienced a significant dynamical interaction (or several such
interactions) in order to eject it from the GC and place it on its current orbit.
Based on an analysis of stellar velocities for stars within $r_{2d} \lesssim$ 20'',
\citet{Trippe:2008sf} argue that IRS 9 does not exhibit an excessively high
velocity as several stars in their sample
exhibit $v3d >$ 358 km s$^{-1}$.
However, their sample spans a range of radii,
and since the velocity dispersion of the NSC increases with decreasing
radius \citep[e.g.][]{Trippe:2008sf, Chatzopoulos:2015lq},
then stars with smaller radii then IRS 9 can be expected to
exhibit higher velocities.
Among the sample of stars examined here with similar radii to IRS 9,
the highest velocity is 339.6 $\pm$ 2.0 km s$^{-1}$, which is
$\sim$30 km s$^{-1}$ slower than the observed velocity of IRS 9 (Figure \ref{fig:vel_dist}).

Whether IRS 9 is a true velocity outlier relative to the NSC is sensitive to
the assumptions made regarding $D(v_{3d}, r)$, especially at the high-end tail
of the distribution.
A detailed dynamical analysis of the NSC is required to
rigorously model $D(v_{3d}, r)$ but is beyond the scope of this paper.

\subsection{Possible Dynamical Mechanisms}
\label{sec:mechanisms}

Given the high density of stars in the GC as well as the presence of
an SMBH, there are several dynamical mechanisms that may have acted upon IRS 9 to place it on its current orbit.
Based on the observed velocity of IRS 9 and the constraints on its orbit,
we explore whether IRS 9 could have been produced by the Hills Mechanism ($\mathsection$\ref{sec:hills}),
binary disruption via supernova ($\mathsection$\ref{sec:sne}), or a close two-body interaction/stellar collision ($\mathsection$\ref{sec:collisions}).

\subsubsection{The Hills Mechanism}
\label{sec:hills}

The Hills mechanism, which describes the disruption of a binary when it passes close enough to an SMBH
that the gravitational tidal force on the binary exceeds its binding energy, has been long identified as
a mechanism to eject stars from galactic nuclei \citep[][see also review by \citet{Brown:2015fv}]{Hills:1988kq}.
This process has been invoked to explain the existence of
hypervelocity stars, which exhibit velocities large enough to escape the Galaxy \citep[e.g.][]{Koposov:2020gz}.
However, it is expected to produce stars at a wide range of velocities depending on the
properties of the initial binary star system \citep[e.g.][]{Kenyon:2008bt, Rossi:2014yv, Generozov:2020vw, Verberne:2025yk}.
If IRS 9 was ejected by the Hills Mechanism within the past 0.4 Myr (the timescale where the orbit
is constrained by the analysis in $\mathsection$\ref{sec:irs9_orb}), then the breakup of its
initial binary is likely to have occurred at the periapse of the orbit at $r$ $\sim$ 0.1 pc.
To determine if this is a viable scenario,
we explore the possible configurations of this initial binary and
evaluate (1) its lifespan in the dense GC environment, and (2) the ejection velocity that would
be achieved by a Hills disruption.

We begin by estimating the age and mass of IRS 9 and its hypothetical binary companion.
As a SiO maser source, IRS 9 is likely an thermally-pulsating Asymptotic Giant Branch star (TPAGB) with a
dynamic circumstellar atmosphere \citep[e.g.][]{Habing:1996pm, Kemball:2007dr}.
The evolved nature of IRS 9 indicates that it is part of the old stellar population
of the NSC.
While the star formation history and metallicity distribution of the NSC is debated,
recent studies suggest that the cluster is dominated by metal-rich stars
([Z] $\sim$ 0.5) with ages between $\sim$5 -- 10 Gyr,
with smaller sub-populations of of metal-poor stars ([Z] $\sim$ -1.0) at similar ages
and/or an intermediate-aged population at $\sim$3 Gyr \citep[e.g.][]{Schodel:2020pp, Chen:2023os}.
Thus, we assume that the initial binary system containing IRS 9 formed
at least $\gtrsim$3 Gyr ago.

Stellar evolution models predict that TPAGB stars in
stellar populations spanning the age and metallicity range found in the NSC
would have masses between 1 M$_{\odot}$ -- 2 M$_{\odot}$ \citep[][]{Choi:2016en}.
We adopt this as the likely mass range for IRS 9.
For the binary companion, we explore two limiting cases:
one case where $M_{irs9}$ / $M_{comp}$ = 1, in which IRS 9 and the companion star are the same mass,
and another where $M_{irs9}$ / $M_{comp}$ = 0.1, in which IRS 9 is the secondary to a massive 10 M$_{\odot}$ -- 20 M$_{\odot}$ primary star.
This spans the range of mass ratios typically observed for binary star systems \citep[0.1 $\leq$ q $\leq$ 1;][]{Sana:2012ez, Moe:2017yg}.

In order to undergo a Hills mechanism disruption at a given radius from SgrA* ($r_{disrupt}$),
the semi-major axis ($a_{bin}$) of the binary system would be:
\begin{equation}
\label{eq:sma}
a_{bin} \sim r_{disrupt} \left(\frac{m_b}{3M_{bh}}\right)^{1/3}\, ,
\end{equation}
where $m_b$ is the total mass of the binary, and $M_{bh}$ is the mass of SgrA* \citep[e.g.][]{Brown:2015fv}.
As before, $M_{bh}$ is assumed to be 4.14 x 10$^6$ M$_{\odot}$.
Assuming $r_{disrupt}$ = 0.1 pc, this calculation yields
162 AU $\lesssim$ $a_{bin}$ $\lesssim$ 204 AU for the $M_{irs9}$ / $M_{comp}$ = 1 system and
286 AU $\lesssim$ $a_{bin}$ $\lesssim$ 360 AU for the $M_{irs9}$ / $M_{comp}$ = 0.1 system.

There are two challenges to this scenario.
First, the large separations mean that the binary lifetime, due to interactions with neighboring stars, is very short.
Second, the resultant velocity of an ejected star due to the Hills mechanism of such a binary is too slow
to produce a star with the velocity observed for IRS 9.
We discuss these two issues below.

\begin{figure*}
\includegraphics[scale=0.35]{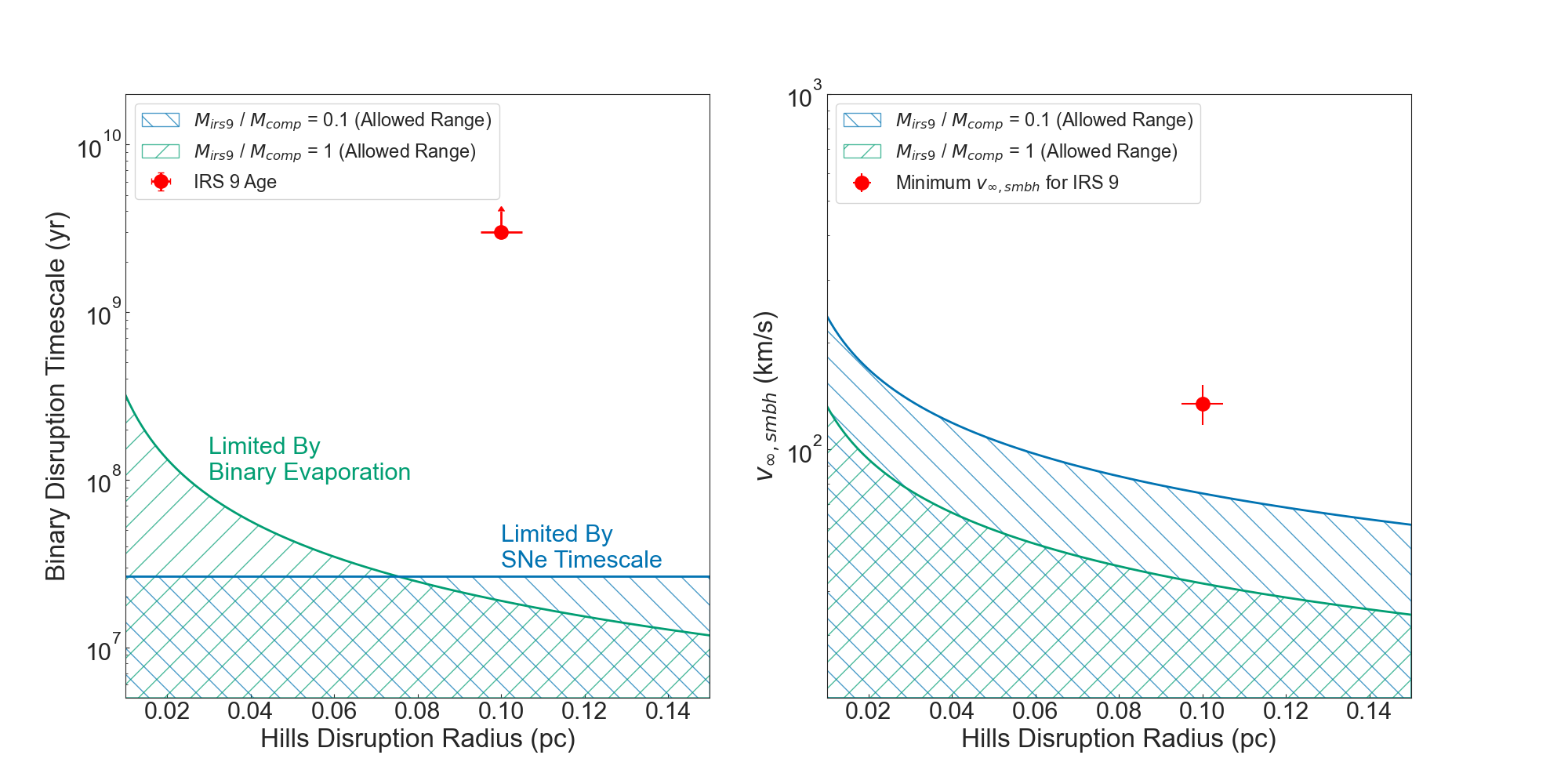}
\caption{Binary disruption timescale (left) and $v_{\infty,smbh}$ (right) as a function of Hills disruption radius
for hypothetical binary systems containing IRS 9.
The allowed ranges for a
$M_{irs9}$ / $M_{comp}$ = 1 system (green hashed region) and a $M_{irs9}$ / $M_{comp}$ = 0.1 system
(blue hashed region) are shown, as well as the current values for IRS 9 (red points).
Assuming that the Hills disruption occurred at the minimum periapse of IRS 9's current orbit ($r_{peri}$ = 0.1 pc),
the disruption timescales of the binary systems are $\lesssim$30 Myr.
Given IRS 9's current age of $\gtrsim$ 3 Gyr, the binary systems could not survive long enough to disrupt within the past 0.4 Myr.
Further, the binary systems cannot achieve the minimum $v_{\infty,smbh}$ of IRS 9 (134 $\pm$ 18 km s$^{-1}$, calculated
assuming $|d_{los}|$ = 0 pc) via the Hills mechanism at a disruption radius of 0.1 pc.
}
\label{fig:disruption}
\end{figure*}

\begin{enumerate}[wide]
\item  \emph{Binary Evaporation Timescale} --- Due to the high stellar density and velocity dispersion at the GC,
binaries are prone to evaporation, a process by which they are gravitationally disrupted
by repeated encounters by passing stars \citep[e.g.][]{Alexander:2014sx}.
Following \citet{Rose:2020pp}, we calculate the orbit-averaged binary evaporation
timescale ($t_{ev}$) for possible IRS 9 binaries (see their Equation 16).
For this calculation, we assume that the original binary system was
on the same highly-eccentric orbit as IRS 9 today ($\mathsection$\ref{sec:irs9_orb}).
This provides an upper limit on $t_{ev}$, since the initial binary system
was presumably on a lower eccentricity orbit than what IRS 9 (as the ejected component) exhibits today,
and $t_{ev}$ generally increases with increasing eccentricity as more time is spent
at larger radii from SgrA*. Under this formulation, we find that the evaporation timescale
of the $M_{irs9}$ / $M_{comp}$ = 1 binary system
with $r_{disrupt}$ = 0.1 pc would be $\lesssim$ 20 Myr.
This is at least 150x smaller than the current age of IRS 9,
and so such a system could not have survived long enough to experience a Hills
mechanism disruption in the past 0.4 Myr (Figure \ref{fig:disruption}).

For a $M_{irs9}$ / $M_{comp}$ = 0.1 binary system, the binary disruption timescale is not set by evaporation
but by the stellar evolution of the primary star.
The MIST stellar evolution models predict that a 10 M$_{\odot}$ star at [Z] = 0.5 (e.g., the massive companion to IRS 9)
will evolve to a supernova in $\sim$30 Myr \citep{Choi:2016en}.
This becomes the upper limit on the disruption timescale of such systems,
independent of the radius of the binary relative to SgrA*.
Therefore, this type of system also could not survived long enough
to disrupt in the past 0.4 Myr (Figure \ref{fig:disruption}).

\item \emph{Ejection Velocity} ---
The velocity of a star ejected via the Hills mechanism (at infinity, relative to the SMBH)
is:
\begin{equation}
\label{eq:hills_vej}
v_{\infty,smbh} = \alpha \sqrt{\frac{2Gm_c}{a_{bin}}} \left(\frac{M_{bh}}{m_b}\right)^{1/6}\, ,
\end{equation}
where $m_c$ is the mass of the captured star and $\alpha$ is a constant
that three-body scattering experiments indicate is on the order of unity \citep[e.g.][]{Sari:2010rn, Rossi:2014yv, Verberne:2025yk}.
The $v_{\infty,smbh}$ of IRS 9
increases as the mass of its hypothetical companion (which would be captured by SgrA* in the Hills scenario) increases.
However, even for a $M_{irs9}$ / $M_{comp}$ = 0.1 system,
$v_{\infty,smbh}$ $\sim$ 75 km s$^{-1}$ at $r_{disrupt}$ = 0.1 pc,
which is
significantly lower than the observed value of $v_{\infty,smbh}$ $\geq$ 134 $\pm$ 18 km s$^{-1}$ (Figure \ref{fig:disruption}).

\end{enumerate}

To summarize, it is highly unlikely that the Hills mechanism could have produced
IRS 9 within the past 0.4 Myr (the timescale covered by the orbit calculations)
due to timescale and ejection velocity arguments.\footnote{We note that this conclusion also holds in the extreme case of a $M_{irs9}$ / $M_{comp}$ = 10 binary system, where IRS 9
has a 0.1 M$_{\odot}$ -- 0.2 M$_{\odot}$ companion. For such a system, the binary evaporation time is even faster ($\lesssim$15 Myr) and the $v_{\infty,smbh}$ from Hills is even lower ($\sim$16 km s$^{-1}$) at $r_{disrupt}$ = 0.1pc.}
To invoke the Hills mechanism, the initial binary containing IRS 9 would
need to have been disrupted within $\sim$30 Myr of its formation
and at $r_{disrupt}$ $<$ 0.1 pc.
Investigating whether such an event could produce a star
with a present-day orbit that is consistent with IRS 9
is beyond the scope of this paper.

\subsubsection{Binary Disruption via Supernova}
\label{sec:sne}
Another dynamical mechanism that could accelerate stars to high velocities
is the disruption of stellar binary
systems due to supernova (SNe). In this case, a companion to a massive star can become unbound due to the sudden
mass loss as well as a natal kick from the SNe itself \citep[e.g.][]{Blaauw:1961ey, Hansen:1997ic, Hobbs:2005al}.
This mechanism is expected to be active at the GC, where it is estimated that only $\lesssim$10\% of massive
binary systems will evaporate before the first SNe occurs \citep{Lu:2019mg, Jurado:2024gt}.
Simulations predict that between 0.5\% \citep{Hoang:2022aq} and 10\% \citep{Bortolas:2017vs}
of such systems within r $\lesssim$0.1 pc from SgrA* will result in the secondary companion
becoming unbound from the SMBH, with that percentage rising to as high as $\sim$20\% when binaries
at larger radii are considered \citep[with most ejections occurring between 0.1 $\leq$ r $\leq$ 0.5 pc;][]{Zubovas:2013jf}.
The ejection velocities of these stars can be quite high \citep[up to $\sim$2000 km s$^{-1}$ for closely bound systems at r = 0.1 pc;][]{Hoang:2022aq}, and so the
observed velocity of IRS 9 ($v_{\infty,smbh}$ $\gtrsim$ 134 $\pm$ 18 km s$^{-1}$) can certainly be achieved.

Similar to the Hills mechanism, the SNe disruption scenario would need to occur
soon after the formation of the initial binary, as the SNe of the massive companion
would be expected to occur within $\lesssim$30 Myr.
Whether IRS 9 could have been ejected early on via this mechanism
and then evolve to its present-day orbit
requires additional analysis that is beyond the scope of this paper.

\subsubsection{Close Two-Body Interactions/Stellar Collisions}
\label{sec:collisions}
Close two-body interactions and stellar collisions are
active dynamical mechanisms near the GC due to the high stellar densities
in the region.
It is predicted that close two-body encounters
could eject stars at velocities similar to what is found for IRS 9
($v_{\infty,smbh}$ $\gtrsim$ 130 km s$^{-1}$)
at a rate between $\sim$10$^{-4}$ yr$^{-1}$ and $\sim$10$^{-5}$ yr$^{-1}$ from the GC \citep{Yu:2003ty}.
Stellar collisions are also thought to be an important dynamical mechanism in
the region, with the ability to modify the stellar population \citep[e.g.][]{Dale:2009an, Mastrobuono-Battisti:2021cn, Balberg:2023ed, Zhang:2023vm, Rose:2023ng, Balberg:2024nn}
and radial density distribution \citep[e.g.][]{Rose:2024vi}.
Recent simulations indicate that such collisions can produce ``unbound''
stars (that is, stars unbound from the SMBH) with $v_{\infty,smbh}$ values similar to IRS 9
due to collisions that occur at R $\sim$ 0.1 pc from SgrA*
\citep{Rose:2025jd}.
The rate of such ejections could also be as high as $\gtrsim$10$^{-5}$ yr$^{-1}$,
depending on the assumptions made regarding energy dissipation during the collisions \citep[e.g.][]{Rose:2025so}.
However, the rate at which these mechanisms might produce
stars with velocities similar to IRS 9 is not well constrained by observations.

\section{Conclusions}
\label{sec:conclusions}
We present proper motion measurements for 23 SiO masers near SgrA* ($r_{2d}$ $<$ 45'', or $r_{2d} \lesssim$ 1.8 pc)
derived from 14 epochs of HST WFC3-IR observations of the GC obtained between 2010 -- 2023.
These measurements are independent of previous radio-based proper motions
of these sources \citepalias{Paine:2022fo, Darling:2023ao} and are often more precise,
by as much as a factor of $\sim$100 for masers at larger radii ($r_{2d}$ $>$ 20").
We combine the HST proper motions with radial velocities for the
sources from the literature in order to calculate their total velocities ($v_{3d}$)
and compare them to the escape velocities at the GC at their projected radii ($v_{esc, max}$).
This comparison reveals that only IRS 9 exhibits $v_{3d}$ $>$ $v_{esc, max}$,
indicating that it must be locally unbound from the GC
($v_{3d}$ = 370 $\pm$ 1.2 km s$^{-1}$, corresponding to $v_{\infty,smbh}$ $\gtrsim$ 134 $\pm$ 18 km s$^{-1}$).
None of the other masers in the sample meet the criteria of being locally unbound,
including previously proposed high-velocity candidates SiO-16, SiO-21, and SiO-25.

While IRS 9 is locally unbound, it is not necessarily destined to escape the GC as a whole.
Assuming that its physical radius is equal to its projected radius of $r_{2d}$ = 0.33 pc (i.e., $d_{los}$ = 0 pc),
we integrate the motion IRS 9 for the past 0.4 Myr in the GC gravitational potential
to place the first constraints on its orbit.
We find that IRS 9 is on a bound orbit with $r_{peri}$ $\geq$ 0.100 $\pm$ 0.005 pc and
$r_{apo}$ $\geq$ 5.25 $\pm$ 0.18 pc, with an eccentricity $e$ $\geq$ 0.91 $\pm$ 0.006
over this time.
We estimate that the probability of observing a star with a velocity
greater than or equal to that of IRS 9 in the NSC at its projected radius
is $\sim1\%$,
identifying it as a candidate to have experienced a
significant dynamical interaction to place it
on its current orbit.

Given the observed velocity of IRS 9 and the constraints
on its orbit, we explore whether the star could have been produced
by the Hills Mechanism.
In this scenario, IRS 9 originated in a binary system that was
tidally disrupted by a dynamical interaction with SgrA*.
Assuming that the disruption occurred in the past 0.4 Myr
(the timescale constrained by the orbit analysis), then the binary was likely
destroyed near the periapse distance of 0.1 pc.
For the range of probable IRS 9 binary systems in this scenario,
we find binary evaporation timescale is too short for the system to
survive long enough to be disrupted in the past 0.4 Myr,
and further, that expected ejection velocity of IRS 9 would be
too slow to explain its velocity today.
Therefore, if IRS 9 were produced by the Hills mechanism then
its initial binary must have been destroyed soon after its
formation ($\lesssim$30 Myr) and at a radius $<$ 0.1 pc.

Alternative dynamical mechanisms at the GC that could
produce stars with velocities similar to that observed for
IRS 9 include binary disruption via SNe, two-body interactions,
and stellar collisions.
If IRS 9 were ejected via the SNe of a massive companion, then the event
also must have occurred within $\sim$30 Myr of its
formation due to the stellar evolution timescale of the companion.
Two-body interactions and stellar collisions
have been predicted to eject stars from the GC with
velocities similar to that of IRS 9, although the rate of such
events is uncertain.
Future studies are required to determine which
(if any) of these mechanisms is most likely to have
acted upon IRS 9.
Moving forward, constraining the population of high-velocity/locally unbound
stars near the GC will allow us to better understand
these various dynamical mechanisms and the rate at which they occur.

\acknowledgements
The authors thank Elena Rossi for valuable discussions regarding possible dynamical mechanisms for IRS 9,
as well as the referee whose comments improved this paper.
M.W.H. is supported by the Brinson Prize Fellowship.
S.C.R. thanks the Lindheimer Postdoctoral Fellowship for support.
S.N. acknowledges the partial support of NSF-BSF grant AST-2206428, as well as Howard and Astrid Preston for their generous support.
A.M.G. acknowledges support from her Lauren B. Leichtman and Arthur E. Levine Endowed Astronomy Chair, as well
as from the Gordon E. \& Betty I. Moore Foundation (award \#11458) and GC Star Society.
This work is based on observations made with the NASA/ESA Hubble Space Telescope, obtained at the Space Telescope Science Institute, which is operated by the Association of Universities for Research in Astronomy, Inc., under NASA contract NAS 5-26555. The observations are associated with programs GO-11671, GO-12318, GO-12667, GO-13049, GO-15199, GO-15498, GO-16004, GO-15894, GO-16681, and GO-16990. This research has made extensive use of the NASA Astrophysical Data System.

\facilities{HST (WFC3-IR)}

\software{AstroPy \citep{Astropy-Collaboration:2013kx},  \texttt{galpy} \citep{Bovy:2015bd}, Matplotlib \citep{Hunter:2007}, numPy \citep{harris2020array}, \texttt{KS2} \citep{Anderson:2008qy}, SciPy \citep{Virtanen:2020fp}}

\clearpage

\bibliography{ms.bib}

\clearpage

\appendix

\section{An HST Proper Motion Catalog of SiO Masers Near the GC}
\label{app:pm_cat}

\begin{figure}
\includegraphics[scale=0.35]{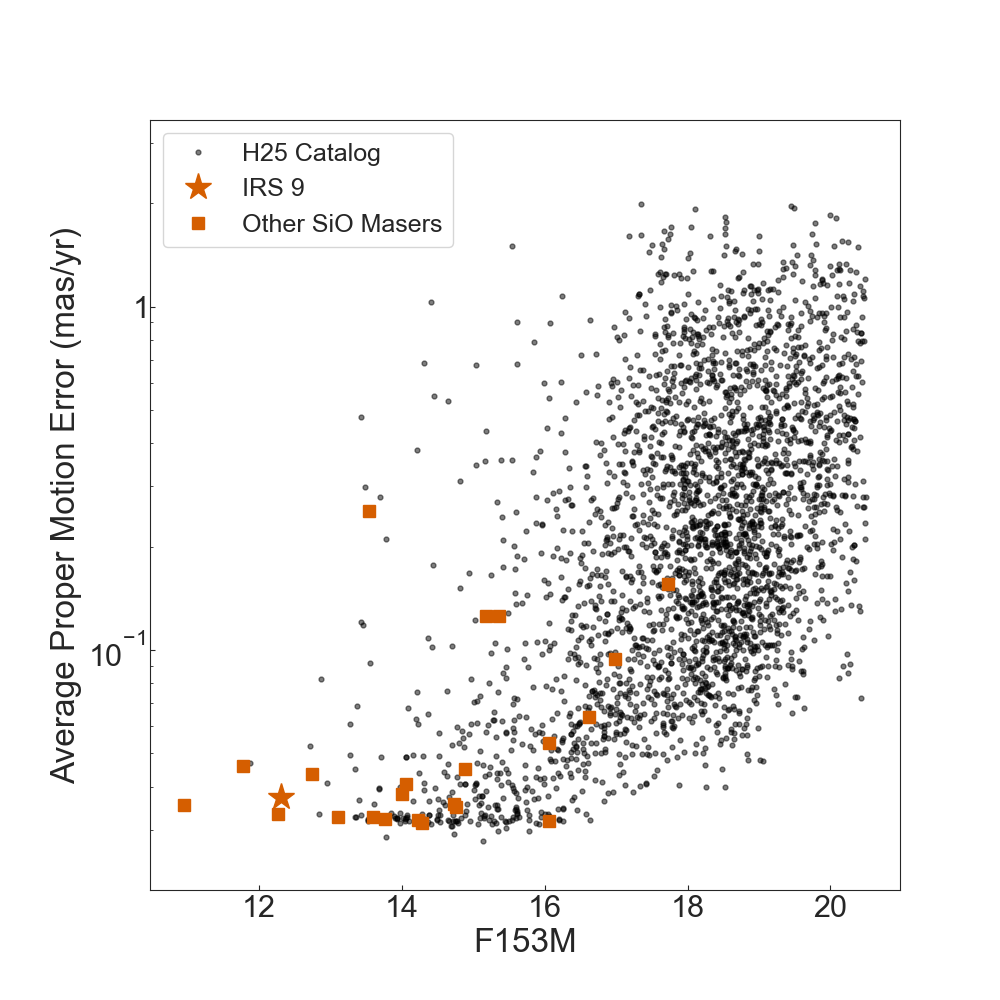}
\caption{HST proper motion error as a function of magnitude for the
SiO masers compared to the HST catalog of astrometric reference stars within R $\leq$25'' from \citetalias{Hosek:2025kf}.
IRS 9 is highlighted as a red star while the other masers are denoted as red squares.
The masers exhibit a median proper motion error of 0.038 mas yr$^{-1}$ across the sample.}
\label{fig:masers_pm_errs}
\end{figure}

The HST proper motions and updated 3D velocities of the
23 SiO masers examined in this study is presented in Table \ref{tab:pm_cat}.
Proper motions are obtained using the same methodology as \citetalias{Hosek:2025kf} (see their $\mathsection$3.2),
which utilizes Gaussian Processes to simultaneously model
systematic errors in the astrometry, if the data requires it.
The parameters of the proper motion fits are reported as:
\begin{equation}
\delta_{\alpha^*}^{s}(t) = \delta_{\alpha^*_0}^{s} + \mu_{\alpha*}^{s} (t - t_0)
\end{equation}
\begin{equation}
\delta_{\delta}^{s}(t) = \delta_{\delta_0}^{s} + \mu_{\delta}^{s} ( t - t_0)\, ,
\end{equation}
where ($\delta_{\alpha^*_0}^{s}$, $\delta_{\delta_0}^{s}$) is the projected
position of the star relative to SgrA* at time $t_0$ and ($\mu_{\alpha*}^{s}$,
$\mu_{\delta}^{s}$) is the proper motion of the star relative to SgrA*.
The proper motion errors for the masers are generally consistent with those
obtained for stars at similar magnitudes extracted from this dataset (Figure \ref{fig:masers_pm_errs}, \citetalias{Hosek:2025kf}).
Additional information provided in Table \ref{tab:pm_cat} includes the 3D velocities
of the masers ($\mathsection$\ref{sec:maser_pms_text}), the GC escape velocities at their projected
radii ($\mathsection$\ref{sec:masers_3d}), and their average HST photometry (F127M, F139M, and F153M filters).

\begin{rotatetable}
\movetableright=1mm
\begin{deluxetable*}{l l c c c c c c c c c c c c c c c c c }
\tablewidth{0pt}
\tabletypesize{\tiny}
\tablecaption{HST Proper Motion Catalog of SiO Masers}
\tablehead{
\colhead{HST ID} & \colhead{Name} & \colhead{$r_{2d}$} & \colhead{F153M} &\colhead{$\delta_{\alpha^*_0}^{s}$} & \colhead{$\sigma_{\delta_{\alpha^*_0}^{s}}$} &
\colhead{$\delta_{\delta_0}^{s}$} & \colhead{$\sigma_{\delta_{\delta_0}^{s}}$} &
\colhead{$\mu_{\alpha*}^{s}$} & \colhead{$\sigma_{\mu_{\alpha*}^{s}}$} & \colhead{$\mu_{\delta}^{s}$} & \colhead{$\sigma_{\mu_{\delta}^{s}}$} &
\colhead{$t_0^{s}$} & \colhead{$v_{3d}$} & \colhead{$\sigma_{v_{3d}}$} & \colhead{$v_{esc, max}$} & \colhead{$\sigma_{v_{esc, max}}$} &
\colhead{Model} & \colhead{Ref} \\
&   & (pc) & (mag)  & ($''$) & (mas) & ($''$) & (mas) & (mas/yr) & (mas/yr) & (mas/yr) & (mas/yr) & (yr) & (km/s) & (km/s) & (km/s) & (km/s) & &
}
\startdata
HST\_NSC\_006749 & IRS9 & 0.33 & 12.3 & 5.7087 & 0.24 & -6.3081 & 0.54 & 2.95 & 0.03 & 2.33 & 0.03 & 2018.0060 & 370.4 & 1.2 & 331.1 & 6.5 & 2 & 1 \\
HST\_NSC\_013716 & IRS15NE & 0.44 & 13.5 & 1.1823 & 1.71 & 11.1953 & 1.32 & -1.17 & 0.29 & -6.53 & 0.22 & 2019.9712 & 255.1 & 8.4 & 290.4 & 6.2 & 5 & 1 \\
HST\_NSC\_006721 & IRS28 & 0.47 & 13.6 & 10.4958 & 0.24 & -5.8933 & 0.52 & 1.29 & 0.02 & -5.52 & 0.02 & 2018.2976 & 224.2 & 1.1 & 281.7 & 6.1 & 0 & 1 \\
HST\_NSC\_013570 & IRS7 & 0.22 & 11.0 & 0.0325 & 0.25 & 5.4753 & 0.53 & -0.13 & 0.03 & -4.68 & 0.03 & 2016.4545 & 212.8 & 3.9 & 409.9 & 7.8 & 0 & 1 \\
HST\_NSC\_001500 & IRS14NE & 0.32 & 13.1 & 0.9500 & 0.24 & -8.1693 & 0.52 & 3.72 & 0.02 & -3.24 & 0.02 & 2018.2453 & 190.0 & 1.3 & 336.5 & 6.6 & 0 & 2 \\
HST\_NSC\_000182 & SiO-18 & 1.26 & 14.8 & -18.6992 & 0.26 & -26.0982 & 0.52 & -4.22 & 0.03 & 0.86 & 0.02 & 2018.2458 & 183.0 & 1.6 & 194.5 & 10.2 & 2 & 2 \\
HST\_NSC\_026835 & SiO-24 & 0.70 & 15.2 & 17.1970 & 0.26 & -4.8063 & 0.54 & 1.04 & 0.12 & -3.71 & 0.13 & 2012.3660 & 149.2 & 5.0 & 237.7 & 6.8 & -1 & 2 \\
HST\_NSC\_033541 & SiO-28 & 1.67 & 16.1 & -1.1102 & 0.28 & -42.5594 & 0.58 & 0.03 & 0.04 & 2.77 & 0.06 & 2018.5707 & 149.0 & 1.8 & 182.3 & 12.2 & 2 & 2 \\
HST\_NSC\_010801 & SiO-11 & 1.58 & 17.7 & 1.7632 & 0.67 & 40.3025 & 0.82 & 2.12 & 0.15 & 2.61 & 0.15 & 2015.7331 & 147.2 & 5.3 & 184.3 & 11.8 & -1 & 1 \\
HST\_NSC\_005236 & SiO-6 & 1.84 & 11.8 & 35.2771 & 0.26 & 30.7005 & 0.56 & 2.57 & 0.03 & 2.46 & 0.05 & 2018.2828 & 146.2 & 1.7 & 179.3 & 12.9 & 2 & 1 \\
HST\_NSC\_029882 & SiO-14 & 1.16 & 14.1 & -7.6216 & 0.28 & -28.4695 & 0.53 & 2.01 & 0.04 & -0.84 & 0.03 & 2018.3923 & 139.0 & 1.1 & 199.3 & 9.6 & 2 & 1 \\
HST\_NSC\_015882 & SiO-25 & 1.48 & 17.0 & -33.0902 & 0.70 & -17.9272 & 0.58 & -1.58 & 0.13 & -0.92 & 0.06 & 2016.5510 & 137.7 & 2.3 & 187.0 & 11.3 & 2 & 2 \\
HST\_NSC\_001403 & IRS12N & 0.30 & 13.8 & -3.2758 & 0.24 & -6.9430 & 0.52 & -1.11 & 0.02 & -2.88 & 0.02 & 2018.0419 & 135.1 & 1.1 & 347.8 & 6.8 & 0 & 1 \\
HST\_NSC\_029099 & SiO-19 & 1.06 & 14.2 & 16.2528 & 0.25 & -21.6642 & 0.53 & 2.70 & 0.03 & 1.23 & 0.02 & 2017.3928 & 117.6 & 1.3 & 204.4 & 9.0 & 0 & 2 \\
HST\_NSC\_015623 & SiO-15 & 0.65 & 14.3 & -12.4686 & 0.23 & -11.0668 & 0.52 & -2.50 & 0.02 & 0.61 & 0.02 & 2018.2361 & 105.2 & 1.3 & 244.5 & 6.6 & 0 & 1 \\
HST\_NSC\_041830 & IRS19NW & 0.92 & 15.4 & 14.5725 & 0.62 & -18.4733 & 0.80 & 1.16 & 0.12 & -0.56 & 0.13 & 2018.3295 & 97.9 & 2.4 & 214.2 & 8.1 & 2 & 1 \\
HST\_NSC\_055002 & SiO-27 & 1.54 & 16.1 & -19.9358 & 0.23 & 33.6746 & 0.52 & -1.26 & 0.02 & 1.76 & 0.02 & 2017.4084 & 94.2 & 1.1 & 185.4 & 11.6 & 0 & 2 \\
HST\_NSC\_006677 & IRS10EE & 0.34 & 16.6 & 7.6851 & 0.32 & 4.1758 & 0.56 & -0.13 & 0.06 & -2.31 & 0.06 & 2018.1952 & 92.9 & 2.3 & 326.7 & 6.5 & 0 & 1 \\
HST\_NSC\_006624 & IRS17 & 0.56 & 12.3 & 13.1282 & 0.24 & 5.5455 & 0.52 & -1.08 & 0.03 & -0.93 & 0.02 & 2018.4157 & 92.3 & 1.0 & 261.3 & 6.3 & 0 & 1 \\
HST\_NSC\_057116 & SiO-20 & 0.97 & 14.9 & -13.8596 & 0.29 & 20.3505 & 0.54 & 0.53 & 0.04 & -1.81 & 0.04 & 2018.4570 & 74.3 & 1.6 & 210.9 & 8.4 & -1 & 2 \\
HST\_NSC\_000852 & SiO-16 & 1.71 & 14.7 & -26.4149 & 0.22 & -34.4759 & 0.54 & 0.17 & 0.02 & -1.72 & 0.04 & 2018.4280 & 66.9 & 1.5 & 181.6 & 12.4 & 2 & 1 \\
HST\_NSC\_027700 & SiO-22 & 1.73 & 12.7 & 41.4090 & 0.26 & 15.1808 & 0.55 & 0.23 & 0.03 & 1.22 & 0.04 & 2018.2940 & 58.7 & 1.5 & 181.1 & 12.5 & 2 & 2 \\
HST\_NSC\_018669 & SiO-21 & 1.82 & 14.0 & 40.9060 & 0.26 & -22.0445 & 0.53 & 0.23 & 0.03 & 0.67 & 0.03 & 2018.4593 & 30.5 & 1.2 & 179.5 & 12.8 & 0 & 2 \\
\enddata
\label{tab:pm_cat}
\tablecomments{Description of columns: \emph{HST ID}: HST Catalog Name,
\emph{Name}: Maser Name,
\emph{$r_{proj}$}: Projected radius of star from SgrA*,
\emph{F153M}: Average F153M mag,
\emph{$\delta_{\alpha^*_0}^{s}$, $\delta_{\delta_0}^{s}$}: star position at $t_0^{s}$ relative to SgrA*,
\emph{$\sigma_{\delta_{\alpha^*_0}^{s}}$, $\sigma_{\delta_{\delta_0}^{s}}$}: error in star position at $t_0^{s}$,
\emph{$\mu_{\alpha*}^{s}$, $\mu_{\delta}^s$}: star proper motion relative to SgrA*,
\emph{$\sigma_{\mu_{\alpha*}^{s}}$, $\sigma_{\mu_{\delta}^{s}}$}: error star proper motion,
\emph{$t_0^{s}$}: reference epoch of proper motion fit,
\emph{$v_{3d}$, $\sigma_{v_{3d}}$}: Value and uncertainty of 3D velocity relative to SgrA*,
\emph{$v_{esc, max}$, $\sigma_{v_{esc, max}}$}: Value and uncertainty of $v_{esc}$ at $r_{2d}$
\emph{Model}: HST kinematic model used -- 0: \emph{poly-only}, 1: \emph{poly+conf}, 2: \emph{poly+sqexp}, 3: \emph{poly+add}, 5: \emph{poly+step} (see \citetalias{Hosek:2025kf}),
\emph{Ref}: Radio proper motion reference -- 1:  \citetalias{Darling:2023ao}, 2: \citetalias{Paine:2022fo}
}
\tablecomments{The table is also provided in machine-readable format.}
\end{deluxetable*}
\end{rotatetable}

\section{Comparing the HST and Radio Proper Motion Measurements of the Masers}
\label{sec:maser_comp}

The radio proper motion measurements adopted for the 23 masers are taken from \citetalias{Darling:2023ao}
and \citetalias{Paine:2022fo}.
Of the sample, 12 of the sources originate from \citetalias{Darling:2023ao}, which defines the most
recent radio-based astrometric reference frame for the GC.
These sources are primarily close to SgrA* ($r_{2d}$ $\lesssim$ 20'') and are among the best-measured
sources, with a history of radio astrometric measurements that extends over a $\sim$26 year time baseline \citep{Reid:2003ai, Reid:2007jk}.
The radio measurements for the remaining 11 sources comes from \citetalias{Paine:2022fo},
whose observations extend over a larger field and thus the sources are mostly at larger
radii ($r_{2d}$ $\gtrsim$ 20''). These sources generally have fewer radio-based astrometric measurements.
The radio reference used for each maser is noted in Table \ref{tab:pm_cat}.

\begin{figure*}
\includegraphics[scale=0.35]{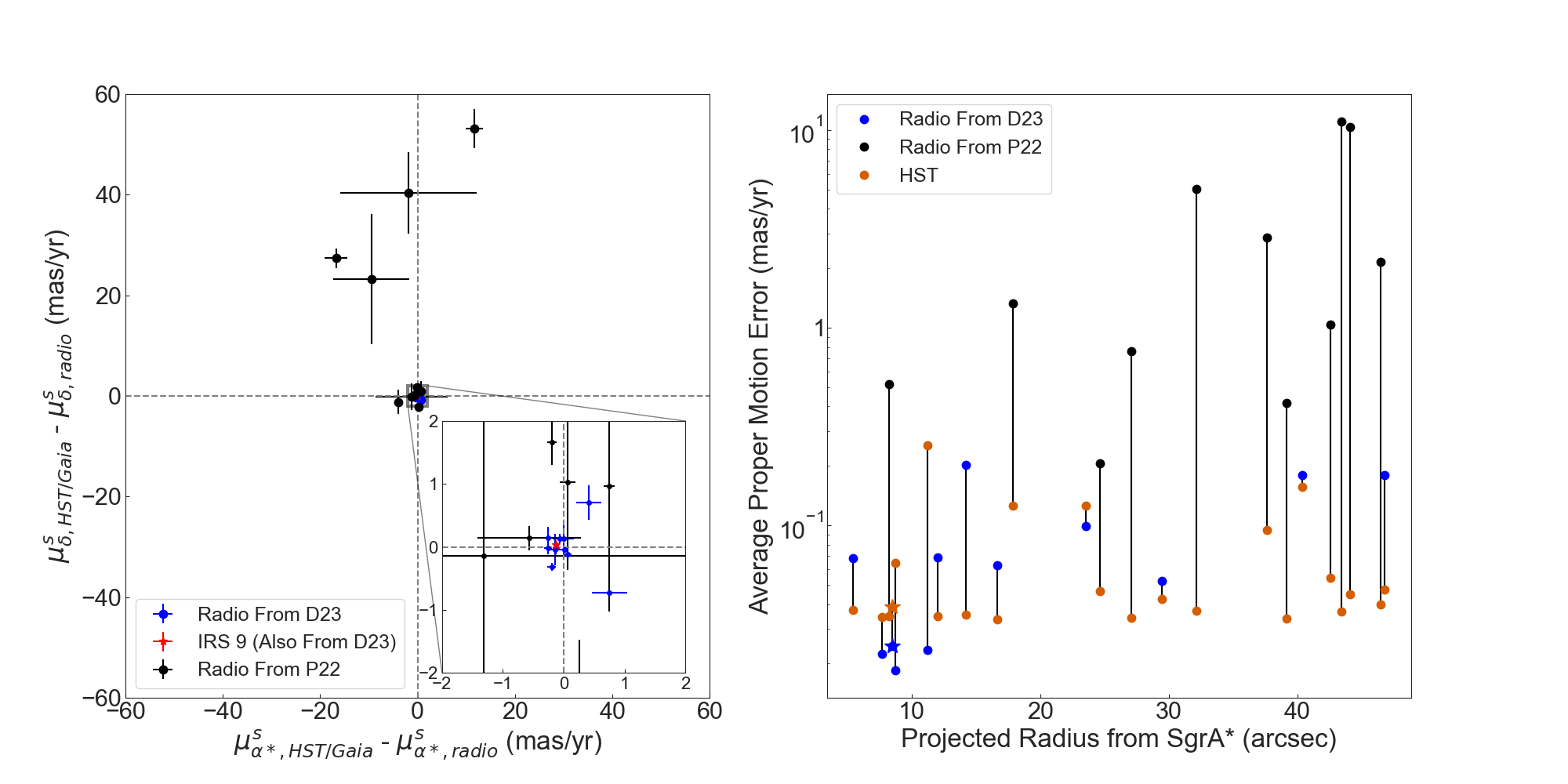}
\caption{
A comparison of the HST and radio proper motion measurements for the masers (left panel) and their corresponding uncertainties (right panel).
Left: The measurements show good agreement for the masers with radio measurements from \citetalias{Darling:2023ao} (blue points),
but show larger discrepancies for those from \citetalias{Paine:2022fo} (black points).
Particularly large discrepancies ($\gtrsim$ 20 mas yr$^{-1}$) are found for SiO-16, SiO-21, and SiO-25.
Right: The HST proper motion errors (red points) are lower for many sources, especially for those at larger distances
where less radio astrometry is available (R $\gtrsim$ 20"). Vertical black lines connect the HST and radio proper motion errors
for the same source. In both panels, IRS 9 is highlighted with a star symbol.
}
\label{fig:masers_comp}
\end{figure*}

A comparison between the HST and radio proper motion measurements
is shown in the left panel of Figure \ref{fig:masers_comp}.
The agreement between the measurements for the \citetalias{Darling:2023ao} masers
is quite good. As discussed in detail by \citetalias{Hosek:2025kf},
the HST and radio proper motion measurements for the \citetalias{Darling:2023ao} sources are consistent
to within 0.041 mas yr$^{-1}$ at 99.7\% confidence.
IRS 9 is a part of this sample.
However, much larger discrepancies are found for many of the
\citetalias{Paine:2022fo} masers, most notably for the proposed high-velocity sources
SiO-16, SiO-21, and SiO-25, which show discrepancies of $\gtrsim$ 20 mas yr$^{-1}$.
As noted by \citetalias{Paine:2022fo}, this is likely due to the limited radio astrometry
available for these sources, which were only measured in 2 radio epochs spanning a
$\sim$2 year time baseline.

The HST and radio proper motion uncertainties are compared in the right
panel of Figure \ref{fig:masers_comp}.
Overall, the HST measurements are more precise for 17 of the masers,
and the median HST proper motion error across the sample is $\sim$4.5x smaller
than the median radio proper motion error (0.038 mas yr$^{-1}$ vs. 0.18 mas yr$^{-1}$).
The largest improvements are achieved for masers at larger
radii from SgrA* ($r_{2d}$ $\gtrsim$ 20"), where the HST uncertainties can be
$\gtrsim$100x smaller than the radio uncertainties.
For sources closer to SgrA* ($r_{2d}$ $\lesssim$ 20"),
the HST and radio measurement uncertainties are more comparable,
with the median proper motion error across the sample being only
$\sim$1.6x smaller for HST compared to radio
(0.03 mas yr$^{-1}$ vs. 0.05 mas yr$^{-1}$).

\section{Adopting Different Line-of-Sight Distances for IRS 9}
\label{app:los_dist}

Throughout this paper we assume that $d_{los}$, the
current line-of-sight distance of IRS 9 relative to SgrA*,
is 0 pc.
In this case, the physical radius of IRS 9 from SgrA* is equal to its
projected radius ($r_{2d}$ = 0.33 pc), which represents the smallest possible current radius of the star.
However, $d_{los}$ is not well constrained for IRS 9.
Here we explore how varying $d_{los}$ impacts the results.

\begin{figure}
\includegraphics[width=0.5\textwidth]{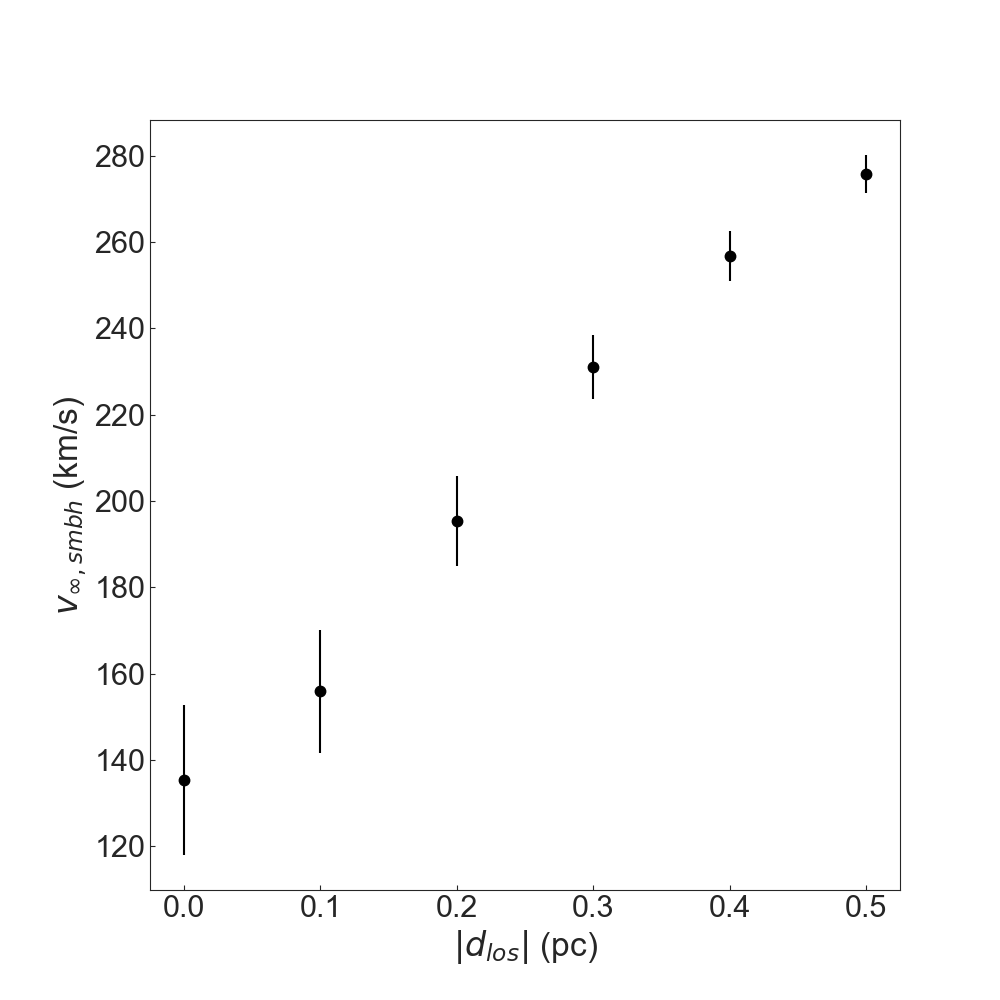}
\includegraphics[width=0.5\textwidth]{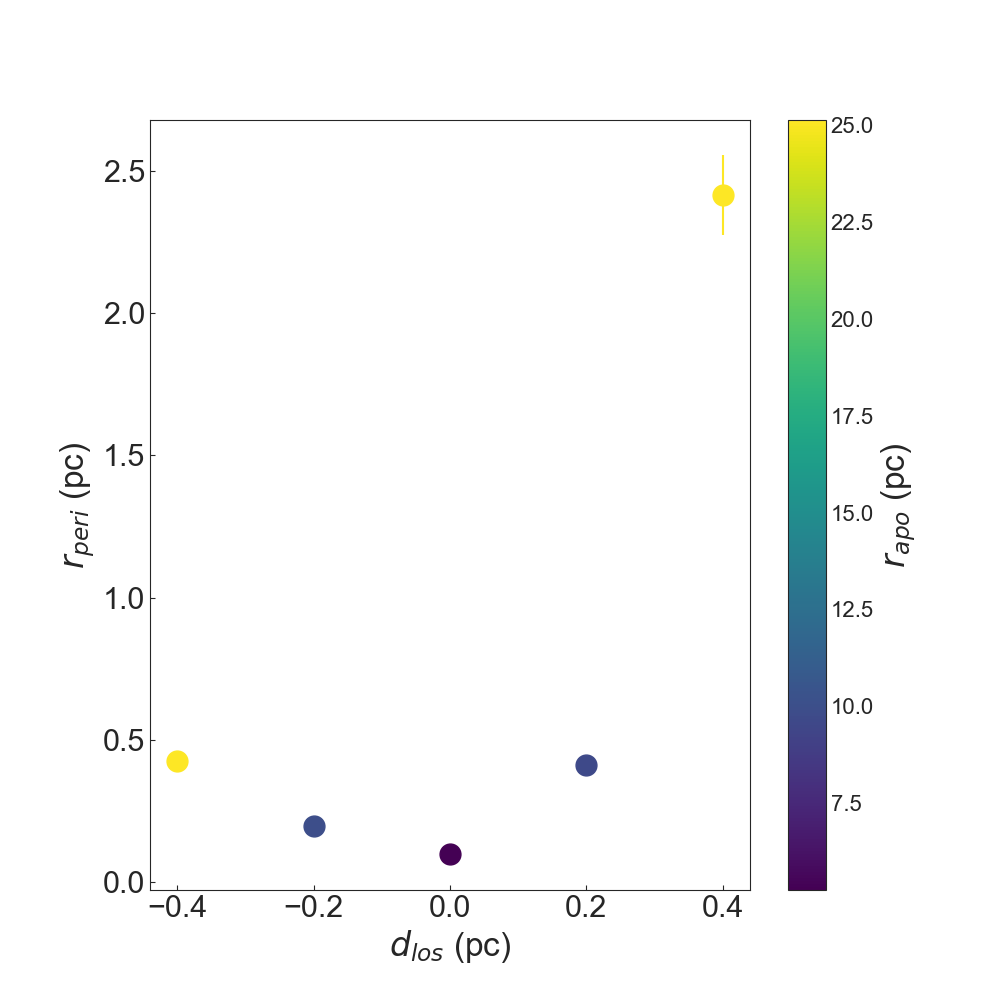}
\caption{
Different calculated properties for the orbit of IRS 9 as a function of $d_{los}$. \emph{Left}: $v_{\infty, smbh}$ increases as $|d_{los}|$ increases, and so the reported value for the $d_{los}$ = 0 pc case ($v_{\infty, smbh}$ = 134 $\pm$ 18 km s$^{-1}$) represents a lower limit. \emph{Right}: $r_{peri}$ as a function of $d_{los}$, with the color of each point corresponding to r$_{apo}$. The minimum values for $r_{peri}$ and $r_{apo}$ are obtained for the d$_{los}$ = 0 pc case, which represents the ``tightest" possible orbit for IRS 9 in the GC.
}
\label{fig:irs9_dlos}
\end{figure}

In $\mathsection$\ref{sec:masers_3d}, we calculate
that $v_{\infty, smbh}$ = 134 $\pm$ 18 km s$^{-1}$ for IRS 9
in the case where $d_{los}$ = 0 pc.
If we instead assume a non-zero value for $d_{los}$,
then the current radius of IRS 9 increases,
and so the gravitational potential energy (due to SgrA*) becomes less negative.
Thus $v_{\infty, smbh}$ increases, for example reaching
a value of 276.2 $\pm$ 4.4 km s$^{-1}$ for $|d_{los}|$ = 0.5 pc (Figure \ref{fig:irs9_dlos}, left panel).

We repeat the orbit simulations in $\mathsection$\ref{sec:irs9_orb}
for $d_{los}$ values of -0.4 pc, -0.2 pc, 0.2 pc, and 0.4 pc.
Since the physical radius of IRS 9 is larger in these cases than when
$d_{los}$ = 0 pc, the periods of the resulting orbits are also larger.
For each case, the orbit integrations are calculated for enough time
to encompass at least two full orbital periods into the past.
For each $d_{los}$ value, the $r_{peri}$ and $r_{apo}$ values derived
are larger than the ones obtained when $d_{los}$ = 0 pc (Figure \ref{fig:irs9_dlos}, middle panel).
This shows that the $r_{peri}$ and $r_{apo}$ values derived for the
$d_{los}$ = 0 pc orbit discussed in the main text of this paper represent
lower limits for IRS 9.

\end{document}